# Photometric and spectroscopic observations of the 2014 eclipse of the complex binary EE Cephei

David Boyd
*Variable Star Section, British Astronomical Association, [davidboyd@orion.me.uk]*

**Abstract**

We report photometric and spectroscopic observations of the 2014 eclipse of EE Cep. This proved to be one of the shallower eclipses on record rather than one of the deepest as predicted. The general shape of the eclipse light curve was similar to that of the 2003 eclipse. The spectral type before and after eclipse was consistent with B5III and became slightly later at mid-eclipse. Total flux in the Hα emission line remained relatively constant through the eclipse.

**Previous observations of EE Cep**

First detected as variable by Romano in 1952 (1), EE Cep was tentatively identified as an eclipsing binary by Romano & Perissinotto (2) who recorded a deep minimum in July 1958. Barbier et al (3), in their study of galactic structure, assigned EE Cep (their star no 74) a spectral type of B6III and a colour excess E(B-V) of 0.52. Meinunger (4) calculated an orbital period of 2050 days (=5.62 years) and deduced from colour indices that the spectral type was B5II-III (incorrectly quoted as B3II-III in reference 5). Only primary eclipses have been seen and each eclipse that has been observed has had a different profile. The depth of the five eclipses between 1958 and 1980 varied between 0.6 and 2.0 magnitudes (4, 5, 6). Spectroscopy reported in (6) gave a spectral type of B5IIIe with broad hydrogen Balmer lines in absorption and Hα and Hβ showing emission cores. No evidence of the companion object was seen in either light curves or spectra.

Mikolajewski & Graczyk (7) suggested a model of EE Cep in which the secondary is a low luminosity object surrounded by an eccentric, slightly tilted, thick disc seen almost edge-on. Precession of this disc would then explain the different eclipse profiles observed at each eclipse. They likened this to the case of ε Aurigae. From measurements outside eclipse, they estimated the absolute luminosity of the primary as Mv=-3.1, its radius as about 10 solar radii and its distance as 2.75 kpc. From measurements of colour indices, they deduced a spectral class of B5III for the primary with an effective temperature of 14,300K.

Analysis of photometric and spectroscopic observations of the 2003 and 2008/9 eclipses reported by Galan et al. (8) confirmed that the main component of the system is a rapidly-rotating Be-type star being eclipsed by a dark, dusty disc around a low luminosity central body, either a single low-mass star or a close binary, which is invisible both photometrically and spectroscopically. The non-central passage of the eclipsing body across the primary produces the observed asymmetrical eclipse profiles. A recently published characterisation of the OGLE LMC-ECL-11893 system as having an eclipsing circumsecondary disk (9) suggests that such systems may not be as rare as previously thought.

According to the ephemeris published by Mikolajewski & Graczyk (7), mid-eclipse was predicted to occur on 2014 August 23 (JD 2456893.44). Galan et al. (10) called for observations of the 2014 eclipse and, based on their analysis of previous eclipses and their precessing dark, dusty disc model, predicted that this would be one of the deepest eclipses reaching about 2 magnitudes.

**Photometric observations of the 2014 eclipse**

V and Ic-band photometry of EE Cep was recorded on 28 nights between July 9 and September 28 using a 0.35m SCT and SXVR-H9 CCD camera. All images were dark subtracted and flat fielded prior to performing aperture photometry. Five comparison stars were used from the AAVSO International Variable Star Database (11). These were combined in a weighted ensemble to reduce the effect of statistical variation in the measurements of individual comparison stars and to monitor each star for possible variability. All measured magnitudes were transformed onto the Johnson-Cousins standard photometric system. Six images were recorded in each waveband on each night and mean magnitudes and standard deviations were calculated for each star for each night.

Table 1 lists the V and Ic magnitudes and uncertainties from the AAVSO Database for the five comparison stars used and their measured mean magnitudes and standard deviations derived from ensemble analysis over all 28 observing runs. The first four of these stars correspond to the stars labelled a, b, c and d in (10). The data show no evidence of variability of any of the comparison stars during this period.

The measured V and Ic magnitudes and (V-Ic) colour indices of EE Cep are listed in Table 2 and plotted in Figure 1. Although there are gaps in coverage due to bad weather and other unavoidable circumstances, the asymmetric profile of the eclipse in both wavebands is clear, as is the colour change during the eclipse. The time of minimum, although not well defined by these data, is consistent with the ephemeris prediction to within a day. The mean magnitudes outside eclipse are V=10.82 and Ic=10.21, and the eclipse depth in V is 0.68 magnitudes and in Ic is 0.58 magnitudes. This is one of the shallower eclipses recorded so the prediction that it would be one of the deepest turned out to be wrong. At mid-eclipse the V-band flux is about 53% of its out-of-eclipse level and the (V-Ic) colour index is 0.1 magnitudes redder.

The profiles of light curves and colour indices are similar to those observed in the 2003 eclipse (8) including the small dip in the V-band light curve at JD 2456862, about 31 days before mid-eclipse, and the small blue "bump" in the (V-Ic) colour index at JD 2456884, about 9 days before mid-eclipse. The out-of-eclipse V magnitude reported here is essentially the same as that reported in (8) for the 2003 and 2008/9 eclipses while the Ic magnitude is approximately 0.2 magnitudes fainter. This can largely be explained by the difference between the Ic magnitudes adopted for the comparison stars in this work and those given in Mikolajewski et al. (12) and used in (8).

**Spectroscopic observations of the 2014 eclipse**

Low resolution (R~1000) spectra were obtained on 17 nights between July 9 and September 28 with a LISA spectrometer and SXVR-H694 CCD camera on a 0.25m SCT. The average spectral resolution was 6Å and SNR about 100 at 6000Å. All spectra were dark-subtracted and flat-fielded and were wavelength calibrated using an internal neon lamp plus Balmer absorption lines in the spectrum of the B8IV star HD 212454 located 2.5˚ from EE Cep. Correction for instrument and atmospheric extinction effects was effected by comparing the recorded spectrum of HD212454 with its spectrum in the MILES Library (13). Finally the spectra of EE Cep were flux calibrated using the method described on the ARAS Forum (14) which uses the V magnitude of the star at the time each spectrum was recorded. As noted in, for example (7), the interstellar reddening E(B-V) in the direction of EE Cep is about 0.5. Spectra were corrected for extinction and reddening using this colour excess and the formulae for normalised extinction Aλ/Av given by Cardelli et al. (15).

The dates and total integration times of spectra are listed in Table 3. Figures 2 and 3 show the flux calibrated spectra of EE Cep taken during eclipse ingress and egress respectively. Spectra before and after the eclipse are consistent with spectral type B5III. The hydrogen Balmer lines are prominent and the spectra also show several weak absorption lines plus the usual atmospheric absorption features. At mid-eclipse the flux is reduced by almost a half and the spectral balance shifts towards red indicating a change to a slightly later spectral type. These changes are consistent with the photometric observations. The change in spectral type bears out the report by Mikolajewski & Graczyk (7) of a change from B5 to B6 during eclipse.

Hα is in emission throughout the eclipse. Table 3 lists and Figure 4 plots the Hα equivalent width, continuum flux at Hα and total flux of the Hα emission line for each spectrum, the latter being the product of equivalent width and continuum flux. The error bars represent estimated uncertainties of 0.5Å in the Hα equivalent width and 2E-14 erg/cm2/s/Å in the continuum flux at Hα. Since the Hα equivalent width and continuum flux plots have similar profiles, both reminiscent of the photometric light curves, the total flux in the Hα emission line remains relatively unchanged through the eclipse. At the observed resolution, the Hβ line is in absorption with an emission core and the Hγ and Hδ lines are in absorption throughout the eclipse.

**Summary**

The light curves of the 2014 eclipse of EE Cep were similar to those recorded in the 2003 eclipse including a small dip in the V-band light curve about 31 days before mid-eclipse and a small blue "bump" in the (V-Ic) colour index about 9 days before mid-eclipse. The prediction that the 2014 eclipse would be deep proved to be wrong as this was one of the shallower eclipses. The spectral type before and after eclipse was consistent with B5III and became slightly later at mid-eclipse. The total flux of the Hα emission line remained relatively unchanged through the eclipse.

**Acknowledgments**

I am grateful to Robert Smith for helpful comments and I acknowledge use of variable star observations from the AAVSO International Database contributed by observers worldwide and used in this research.

| GSC name | AAVSO id | AAVSO V mag | AAVSO Ic mag | Measured V mag | Measured Ic mag |
|---|---|---|---|---|---|
| 3973-1177 | 104 | 10.399 ± 0.007 | 10.028 ± 0.014 | 10.396 ± 0.006 | 10.034 ± 0.005 |
| 3973-2150 | 112 | 11.247 ± 0.006 | 10.986 ± 0.018 | 11.246 ± 0.004 | 10.986 ± 0.005 |
| 3973-1103 | 113 | 11.251 ± 0.006 | 10.907 ± 0.009 | 11.251 ± 0.003 | 10.901 ± 0.004 |
| 3973-1261 | 119 | 11.865 ± 0.009 | 11.496 ± 0.020 | 11.871 ± 0.004 | 11.508 ± 0.008 |
| 3973-1335 | 129 | 12.898 ± 0.006 | 12.159 ± 0.021 | 12.898 ± 0.012 | 12.149 ± 0.010 |

Table 1: V and Ic magnitudes and uncertainties from the AAVSO Database for five comparison stars used and their measured mean magnitudes and standard deviations derived from ensemble analysis over 28 observing runs.

| Date | JD | V mag | Ic mag | (V-Ic) mag |
|---|---|---|---|---|
| 09-Jul-14 | 2456848.4531 | 10.819 ± 0.007 | 10.200 ± 0.010 | 0.619 ± 0.013 |
| 10-Jul-14 | 2456849.4679 | 10.821 ± 0.009 | 10.211 ± 0.006 | 0.610 ± 0.007 |
| 13-Jul-14 | 2456852.4395 | 10.828 ± 0.004 | 10.216 ± 0.007 | 0.611 ± 0.011 |
| 15-Jul-14 | 2456854.4539 | 10.828 ± 0.005 | 10.210 ± 0.008 | 0.619 ± 0.008 |
| 21-Jul-14 | 2456860.4387 | 10.846 ± 0.007 | 10.228 ± 0.007 | 0.618 ± 0.011 |
| 23-Jul-14 | 2456862.4021 | 10.857 ± 0.009 | 10.233 ± 0.007 | 0.623 ± 0.008 |
| 30-Jul-14 | 2456869.3998 | 10.849 ± 0.004 | 10.236 ± 0.013 | 0.613 ± 0.014 |
| 06-Aug-14 | 2456876.3702 | 10.961 ± 0.005 | 10.327 ± 0.009 | 0.634 ± 0.007 |
| 10-Aug-14 | 2456880.4237 | 11.019 ± 0.006 | 10.368 ± 0.008 | 0.651 ± 0.007 |
| 12-Aug-14 | 2456882.3715 | 11.047 ± 0.005 | 10.405 ± 0.014 | 0.643 ± 0.015 |
| 14-Aug-14 | 2456884.4228 | 11.076 ± 0.009 | 10.435 ± 0.007 | 0.641 ± 0.010 |
| 15-Aug-14 | 2456885.3708 | 11.088 ± 0.010 | 10.445 ± 0.013 | 0.643 ± 0.018 |
| 17-Aug-14 | 2456887.3837 | 11.193 ± 0.006 | 10.539 ± 0.004 | 0.654 ± 0.008 |
| 23-Aug-14 | 2456893.3529 | 11.468 ± 0.007 | 10.767 ± 0.009 | 0.701 ± 0.014 |
| 25-Aug-14 | 2456895.4113 | 11.502 ± 0.006 | 10.787 ± 0.007 | 0.715 ± 0.007 |
| 28-Aug-14 | 2456898.3644 | 11.348 ± 0.005 | 10.680 ± 0.005 | 0.668 ± 0.006 |
| 30-Aug-14 | 2456900.4493 | 11.214 ± 0.009 | 10.566 ± 0.011 | 0.649 ± 0.009 |
| 31-Aug-14 | 2456901.3449 | 11.168 ± 0.005 | 10.515 ± 0.004 | 0.653 ± 0.006 |
| 01-Sep-14 | 2456902.4316 | 11.127 ± 0.004 | 10.494 ± 0.006 | 0.633 ± 0.006 |
| 02-Sep-14 | 2456903.4017 | 11.089 ± 0.009 | 10.448 ± 0.015 | 0.641 ± 0.022 |
| 03-Sep-14 | 2456904.4302 | 11.044 ± 0.005 | 10.423 ± 0.009 | 0.620 ± 0.007 |
| 07-Sep-14 | 2456908.3712 | 10.862 ± 0.007 | 10.251 ± 0.008 | 0.611 ± 0.008 |
| 16-Sep-14 | 2456917.3738 | 10.829 ± 0.011 | 10.211 ± 0.014 | 0.618 ± 0.023 |
| 17-Sep-14 | 2456918.3226 | 10.819 ± 0.010 | 10.205 ± 0.012 | 0.613 ± 0.014 |
| 21-Sep-14 | 2456922.3067 | 10.834 ± 0.004 | 10.210 ± 0.007 | 0.624 ± 0.007 |
| 22-Sep-14 | 2456923.4110 | 10.824 ± 0.008 | 10.208 ± 0.011 | 0.616 ± 0.012 |
| 24-Sep-14 | 2456925.4928 | 10.818 ± 0.007 | 10.201 ± 0.007 | 0.616 ± 0.013 |
| 28-Sep-14 | 2456929.3090 | 10.822 ± 0.005 | 10.204 ± 0.004 | 0.619 ± 0.003 |

Table 2: Measured V and Ic magnitudes and (V-Ic) colour indices of EE Cep.

| Date | JD | Integration time (sec) | Hα EW (Å) | Hα continuum flux (erg/cm2/s/Å) | Hα line flux (erg/cm2/s) |
|---|---|---|---|---|---|
| 09-Jul-14 | 2456848.4914 | 3900 | -21.4 | 4.2E-13 | 9.00E-12 |
| 13-Jul-14 | 2456852.4780 | 3600 | -21.6 | 4.3E-13 | 9.27E-12 |
| 15-Jul-14 | 2456854.4527 | 4200 | -21.4 | 4.2E-13 | 9.01E-12 |
| 23-Jul-14 | 2456862.4914 | 3900 | -20.2 | 4.1E-13 | 8.26E-12 |
| 31-Jul-14 | 2456869.5334 | 3600 | -18.9 | 4.0E-13 | 7.56E-12 |
| 06-Aug-14 | 2456876.3990 | 3600 | -22.1 | 3.8E-13 | 8.38E-12 |
| 18-Aug-14 | 2456887.5487 | 3600 | -25.7 | 3.1E-13 | 7.96E-12 |
| 23-Aug-14 | 2456893.4869 | 3600 | -32.6 | 2.6E-13 | 8.47E-12 |
| 25-Aug-14 | 2456895.4587 | 1800 | -33.4 | 2.4E-13 | 8.02E-12 |
| 31-Aug-14 | 2456900.5259 | 4500 | -25.1 | 3.0E-13 | 7.53E-12 |
| 01-Sep-14 | 2456902.4739 | 5700 | -23.1 | 3.2E-13 | 7.39E-12 |
| 02-Sep-14 | 2456903.4199 | 3600 | -22.0 | 3.1E-13 | 6.82E-12 |
| 03-Sep-14 | 2456904.4653 | 5100 | -21.3 | 3.6E-13 | 7.66E-12 |
| 07-Sep-14 | 2456908.3968 | 3600 | -17.3 | 4.1E-13 | 7.10E-12 |
| 16-Sep-14 | 2456917.4009 | 4800 | -19.2 | 4.0E-13 | 7.69E-12 |
| 21-Sep-14 | 2456922.3431 | 3600 | -19.5 | 4.0E-13 | 7.81E-12 |
| 28-Sep-14 | 2456929.3535 | 3900 | -20.4 | 4.2E-13 | 8.58E-12 |

Table 3: Integration time, Hα equivalent width, continuum flux at Hα and total flux of the Hα emission line for each spectrum.

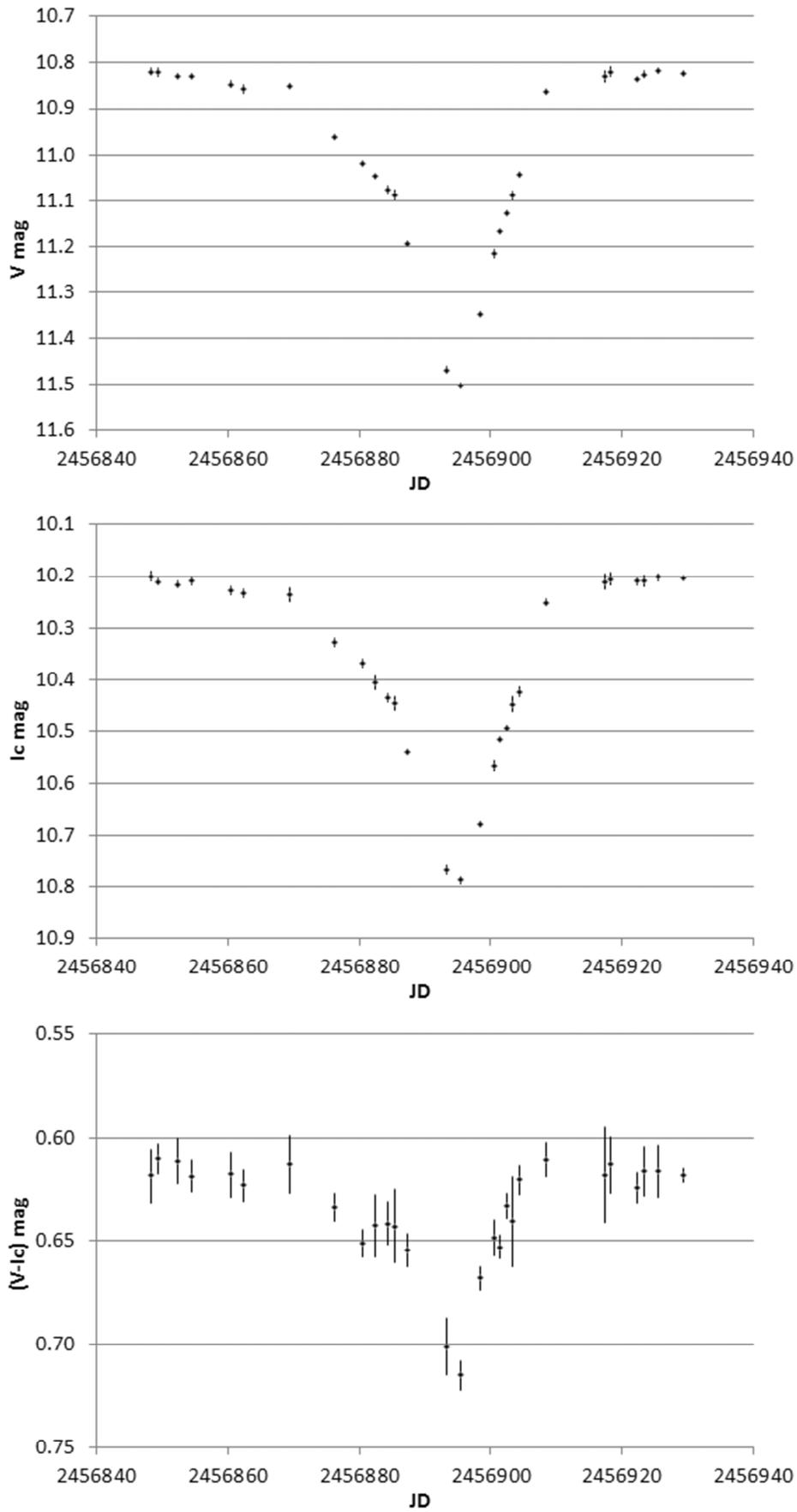

Figure 1: Measured V and Ic magnitudes and (V-Ic) colour indices of EE Cep.

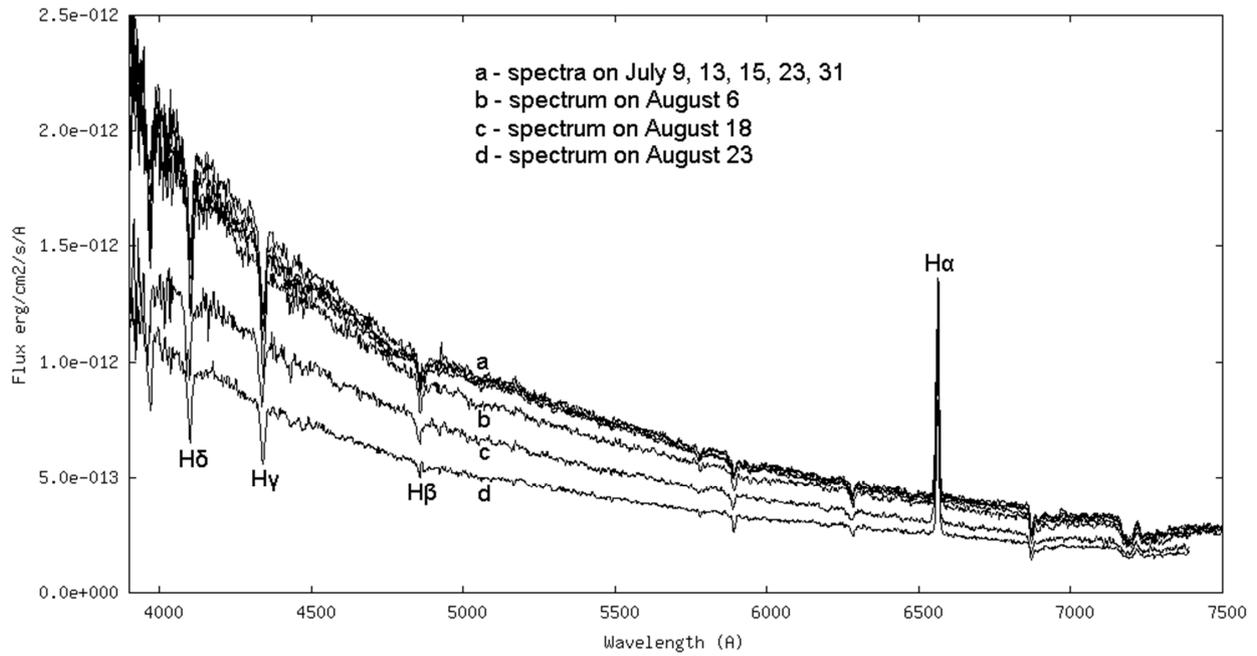

Figure 2: Flux calibrated spectra of EE Cep during eclipse ingress.

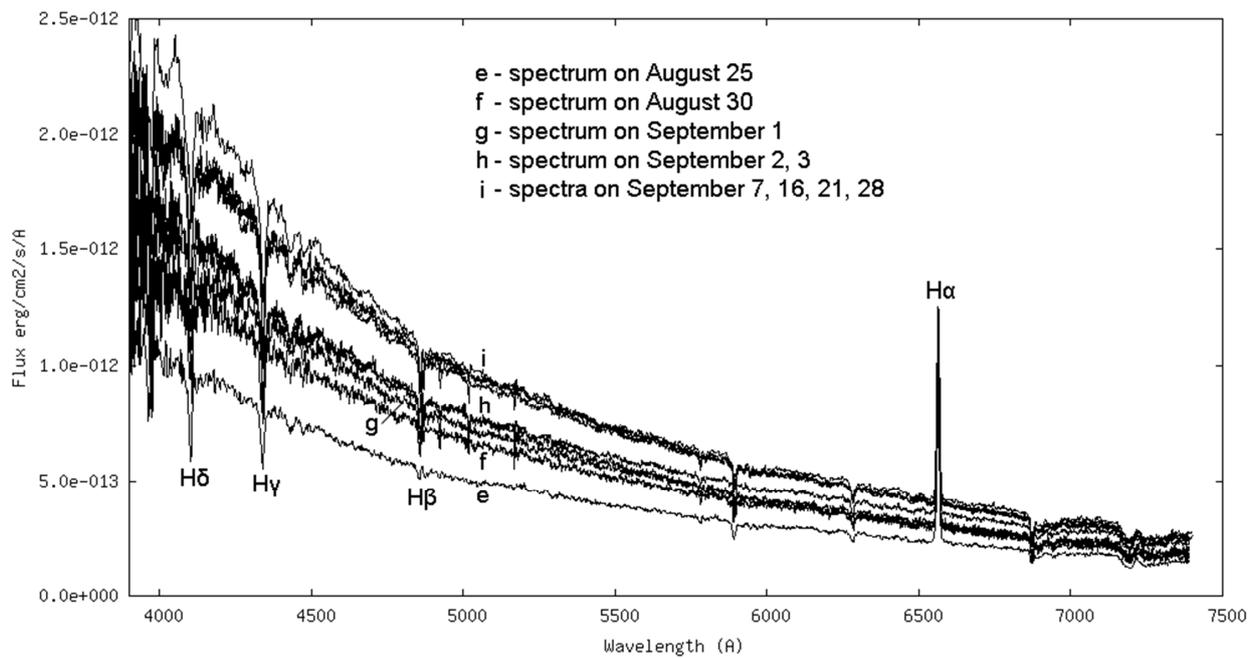

Figure 3: Flux calibrated spectra of EE Cep during eclipse egress.

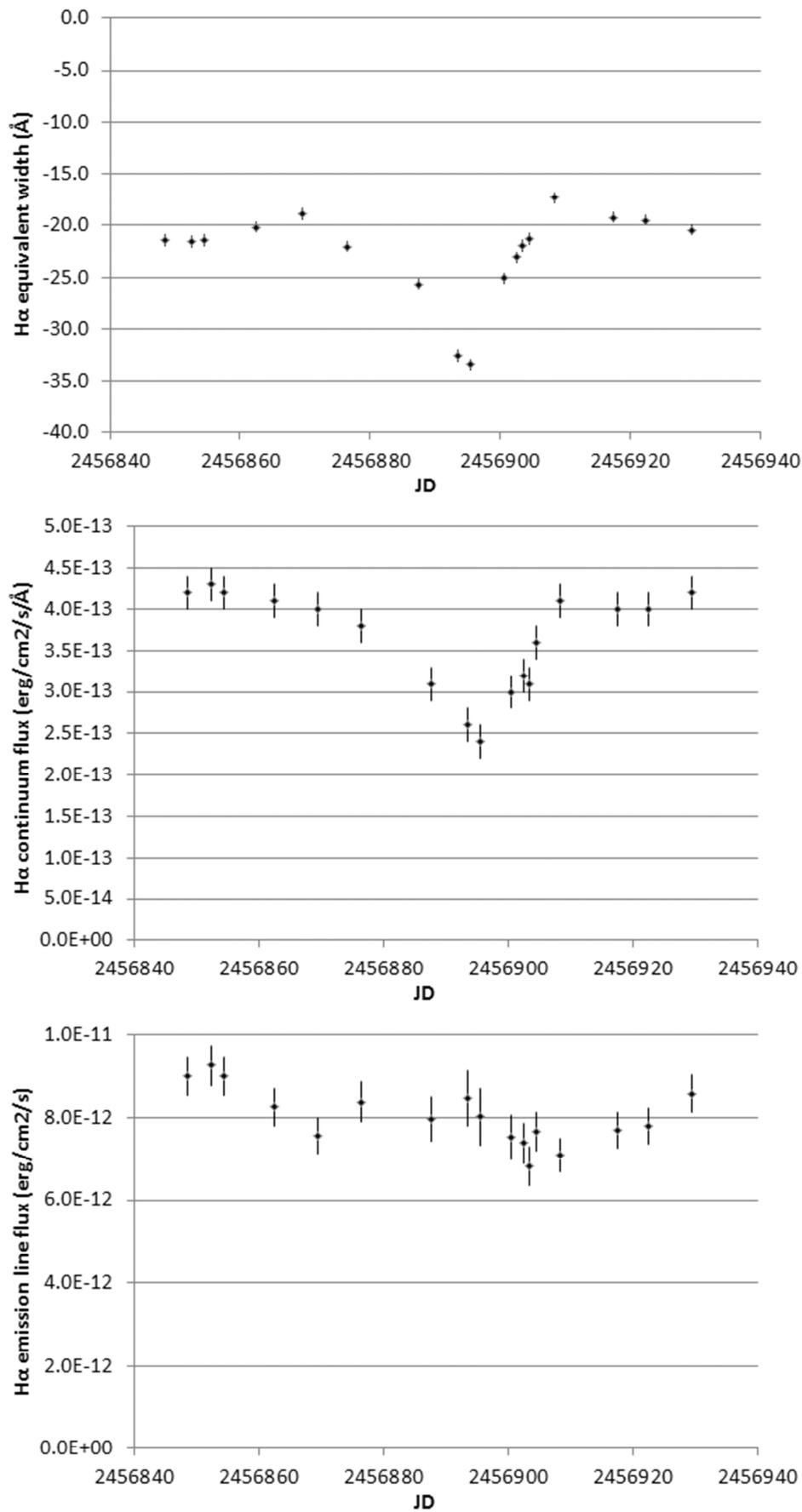

Figure 4: Hα equivalent width, continuum flux at Hα and total flux of the Hα emission line.